# Path Integral Control in Infectious Disease Modeling

Paramahansa Pramanik*
*Assistant Professor, Department of Mathematics and Statistics,University of South Alabama, United States.*

COVID-19, a pandemic that affected the whole world, claimed the lives of almost 1.1 million people in the United States and 7 million worldwide. Prior to the discovery of vaccines, many countries resorted to implementing lock downs to reduce the spread of the virus. Most economies have implemented this policy, except in essential sectors such as public health and safety. Different states in the US have imposed lock downs at different times, based on the severity of the outbreak in their respective regions. Lock downs involve reducing social interactions, leading to a decrease in the transmission of the virus. However, if lock downs remain in effect for too long, people may become hesitant to resume social activities for fear of contracting COVID-19 [3]. Hence, businesses are facing a reduction in the number of consumers and employees, resulting in a decrease in sustainable long-term protability. Furthermore, if a business fails to have enough inventory to with stand the crisis, it may eventually shut down. Since the government is not providing nancial support, it is easy to shut down a business, but dicult to restore it to its original employment levels [3]. It is recommended by the Centers for Disease Control and Prevention (CDC) that anyone infected with Omicron should isolate themselves for five days. This is because a person infected with the virus can spread it to others, so isolation helps reduce transmission. Similarly, if more people are vaccinated, the virus will spread less and fewer people will be affected, thus saving more lives. In their study, Pramanik (2023) [10] determined the best way to decide when to shut down an economy and what rate of vaccination is optimal. They used a healthcare cost function that was minimized while taking into account a stochastic susceptible-infectious-recovered (SIR) dynamic, which was first introduced in Aron et al (1984) [1]. Most models of infectious disease transmission are based on the SIR model. Pramanik's construction can be extended to a generalized random surface to investigate unprecedented shocks, such as the emergence of a new COVID-19 variant, sudden infection due to random interactions caused by travel, and environmental calamities resulting in more exposure to the pandemic. The random surface replaces the jump diffusion of the stochastic differential equations.

In the literature of infectious disease modeling lock-down intensity is defined as the ratio of employment due to COVID-19 to total employment under the absence of it. Therefore, the value of this lock-down intensity lies between 0 and 1 where 0 stands for complete shut-down of an economy. The solution to the optimal "locking down" problem is complex. If an economy imposes strict lockdown policies for a long time, it can effectively reduce the infection rate to a very low level. However, if lockdown policies are relaxed too soon, the infection rate of COVID-19 may increase and reach its peak. Pramanik (2023) [10] assumes that the information regarding the transmission of COVID-19 is incomplete and imperfect, which can lead to multiple Skiba points or multiple solutions. Studies about Skiba points have been rigorously conducted in [4, 11] and [12]. Although there is a growing literature on COVID-19 and its socioeconomic impacts related to extended lockdown time, the length of lockdown and the appropriate time to initiate lockdown have not been studied in depth [3]. Recent studies have investigated the impacts of these components both analytically and empirically in [9, 10].

The Feynman-type path integral approach is based on a social cost minimization problem with a deterministic weight, as explained by Marcet et al. (2019) [6]. This leads to a Wick-rotated Schrödinger-type equation similar to the Hamiltonian-Jacobi-Bellman equation (HJB) [7, 13] and a saddle-point functional equation [6]. Using dynamic programming tools, a numerical algorithm can be developed to find a solution for the infectious disease model. In statistical analysis of infectious disease modeling, crucial factors such as immunity and susceptibility play an important role. It is important to note that the assumption that everyone in a community is susceptible to a pandemic may lead to underestimation of its infectiousness [2]. On the other hand, if people who have previously acquired immunity are able to avoid infection during a pandemic, this suggests high infectiousness. Additionally, the immunity status of individuals, assessed by tests on blood, saliva or excreta samples, is another determinant of the intensity of the pandemic spread [2]. Therefore, random network analysis is essential to determine the spread of infectious diseases.

In tackling generalized non-linear stochastic SIR models, a new approach has been discovered. The traditional HJB approach has led to the curse of dimensionality, making it nearly impossible to obtain an optimal condition through computer simulation. However, a logarithmic transformation can transform a class of non-linear HJB equations into linear equations. This transformation was first used by Schrodinger in the early days of quantum mechanics to relate the HJB equation to the Schrödinger equation [5]. The linear feature of this transformation allows for the replacement of backward integration of the HJB equation over time with the computation of expectation values under a forward diusion process requiring a stochastic integration over trajectories described by a path integral [5, 8]. By solving a particular class of HJB equation locally and providing a stable solution, this approach makes simulation and data handling relatively easier.